\documentclass[]{raa}
\usepackage{graphicx}
\usepackage{times}             
\usepackage{amssymb}
\usepackage{amsmath}
\usepackage{mathrsfs}
\usepackage{float}
\usepackage{natbib}
\usepackage{booktabs}
\usepackage{hyperref}
\usepackage{xspace}

\newcommand{\psrb}{PSR B1259-63\xspace}
\newcommand{\fermi}{$Fermi$-LAT\xspace}

\begin{document}
\title{
The GeV emission of \psrb during its last three periastron passages observed by \fermi
}

\volnopage{Vol.0 (20xx) No.0, 000--000}      
\setcounter{page}{1}          

\author{
Zhi Chang \inst{1}
\and
Shu Zhang \inst{1}
\and
Yu-Peng Chen \inst{1}
\and
Long Ji \inst{2}
\and
Ling-Da Kong \inst{1,3}
\and
Cong-Zhan Liu \inst{1}
}

\institute{
Key Laboratory of Particle Astrophysics,
Institute of High Energy Physics,
Chinese Academy of Sciences,
Beijing 100049, China; $szhang@ihep.ac.cn$\\
\and
Institut f\"ur Astronomie und Astrophysik,
Kepler Center for Astro and Particle Physics,
Eberhard Karls Universit\"at, Sand 1, 72076 T\"ubingen, Germany \\
\and
University of Chinese Academy of Sciences, Beijing 100049, China \\
}

\date{Received~2018 January 29; accepted~2018 July 6}

\abstract{
\psrb is a $\gamma$-ray emitting high mass X-ray binary system,
in which the compact object is a millisecond pulsar.
The system has an orbital period of 1236.7\,d and shows peculiar $\gamma$-ray flares
when the neutron star moves out of the stellar disk of the companion star.
The $\gamma$-ray flare events were firstly discovered by using \fermi 
around the 2010 periastron passage, which was repeated for the 2014 and 2017 periastron passages.
We analyze the \fermi data for all the three periastron passages and
found that in each flare the energy spectrum can be represented well by a simple power law.
The $\gamma$-ray light curves show that in 2010 and 2014 after each periastron 
there are two main flares, 
but in 2017 there are four flares including one precursor about 10\,d after the periastron passage.
The first main flares in 2010 and 2014 are located at around 35\,d after the periastron passage,
and the main flare in 2014 is delayed by roughly 1.7\,d with respect to that in 2010.
In the 2017 flare, the source shows a precursor about 10\,d after the periastron passage,
but the following two flares become weaker and lag behind those in 2010 by roughly 5\,d.
The strongest flares in 2017 occurred 58\,d and 70\,d after the periastron passage.
These results challenge the previous models.
}

\authorrunning{Z. Chang et al.}            
\titlerunning{Three Periastron Passages of PSR B1259-63}  

\maketitle

\section{Introduction}
\label{sect:intro}

PSR B1259-63/LS 2883 is a $\gamma$-ray binary system, 
a member of a special class of X-ray binaries, emitting radiation in  broad wavelength.
Among gamma-ray binaries,
PSR J2032+4127 \citep{abdo2009} and \psrb are the only two for which the nature of the compact object is known:
it consists of a pulsar and a massive main sequence Be star \citep{johnston1992}.
The pulsar PSR B1259-63 is a non-recycled, spin-down powered radio pulsar,
with a spin period of 47.76\,ms and a period derivative of $2.27\times10^{-15}$ \citep{shannon2014}.
The companion star, LS 2883, is a Be star with a mass of $\sim31\,M_{\odot}$.
The binary orbit is highly eccentric, with an eccentricity of $e=0.87$ 
and an orbital period of $P=1236.7$\,d.
The pulsar has a distance of $\sim$0.67\,AU with respect to its companion star at periastron,
which is roughly comparable to the size of the equatorial disk of the companion \citep{johnston1992}.
The pulsar will cross this disk twice in one orbit,
as the orbital plane of the pulsar is thought to be highly inclined
with respect to this equatorial disk \citep{melatos1995}.
Multiwavelength emissions are usually thought to result from the shock interaction between
relativistic pulsar wind and stellar wind.
Around the 2010 and 2014 periastrons of PSR B1259-63, 
enhanced $\gamma$-ray emissions as well as GeV flares were observed
by \fermi at the time when the neutron star moved out of the stellar disk the second time.
\citep{abdo2011, tam2011, caliandro2015}.

Though there are other GeV detected gamma-ray binaries also hosting a Be star 
as a companion (LS\,I\,+61$^\circ$303, \citet{hadasch2012}; HESS J0632+057, \citet{li2017}), 
the GeV emissions observed in PSR B1259-63 during periastron passages are unique.
Here using the latest instrument response functions (IRFs), 
we report the \fermi observations of the 2017 periastron passage of PSR B1259-63, 
and compare it to the 2010 and 2014 passages.
The paper is structured as follows.
The observations and data analysis are described in Section 2,
and the results are provided in Section 3. 
In Section 4 we highlight how these results have constrained the current models.

\section{Observations and data analysis}

The Large Area Telescope (LAT) on board $Fermi$ is an electron-positron
pair production telescope operating at energies from $\sim100$\,MeV to 
greater than $300$\,GeV \citep{atwood2009}.
\fermi observed \psrb during its periastron period in 2010 \citep{abdo2011, tam2011},
2014 \citep{caliandro2015} and 2017 \citep{he2017,johnson2018,tam2018}.

The analysis of \fermi data was performed using the $Fermi$ Science Tools v10r0p5 
package\footnote{\url{https://fermi.gsfc.nasa.gov/ssc/data/analysis/software/}}.
The Pass 8 SOURCE event 
class\footnote{\url{https://fermi.gsfc.nasa.gov/ssc/data/analysis/documentation/Pass8\_usage.html}}
was included in the analysis using the P8R2\_SOURCE\_V6 IRFs.
All $\gamma$-ray photons within an energy range of 100\,MeV--100\,GeV 
and within a circular region of interest with a 10$^\circ$ radius 
centered on PSR B1259-63 were used for this analysis. 
Time intervals, 
when the region around PSR B1259-63 was observed  at a zenith angle less than 90$^\circ$,
were selected to avoid contamination from the Earth limb $\gamma$-rays. 
The Galactic and isotropic diffuse emission components as well as known $\gamma$-ray sources 
within 15$^\circ$ of PSR B1259-63 based on the 3FGL catalog \citep{acero2015} were considered in our analysis. 
The spectral  parameters were fixed to the catalog values, 
except for sources within 3$^\circ$ of PSR B1259-63, for which the flux normalization was left free.
In the modeling of PSR B1259-63, a simple power-law was applied.

The times for three periastron passages of PSR B1259-63 used in our work are 
MJD 55544.693781 (2010-12-14 16:39:02.000 UTC), 
MJD 56781.418307 (2014-05-04 10:02:21.000 UTC) and 
MJD 58018.142833 (2017-09-22 03:25:40.771 UTC).
They are derived from the orbital ephemeris as reported in \cite{shannon2014}.

\section{Results}
\subsection{GeV Flares}

\begin{figure}
  \centering
  \includegraphics[width=0.48\textwidth] {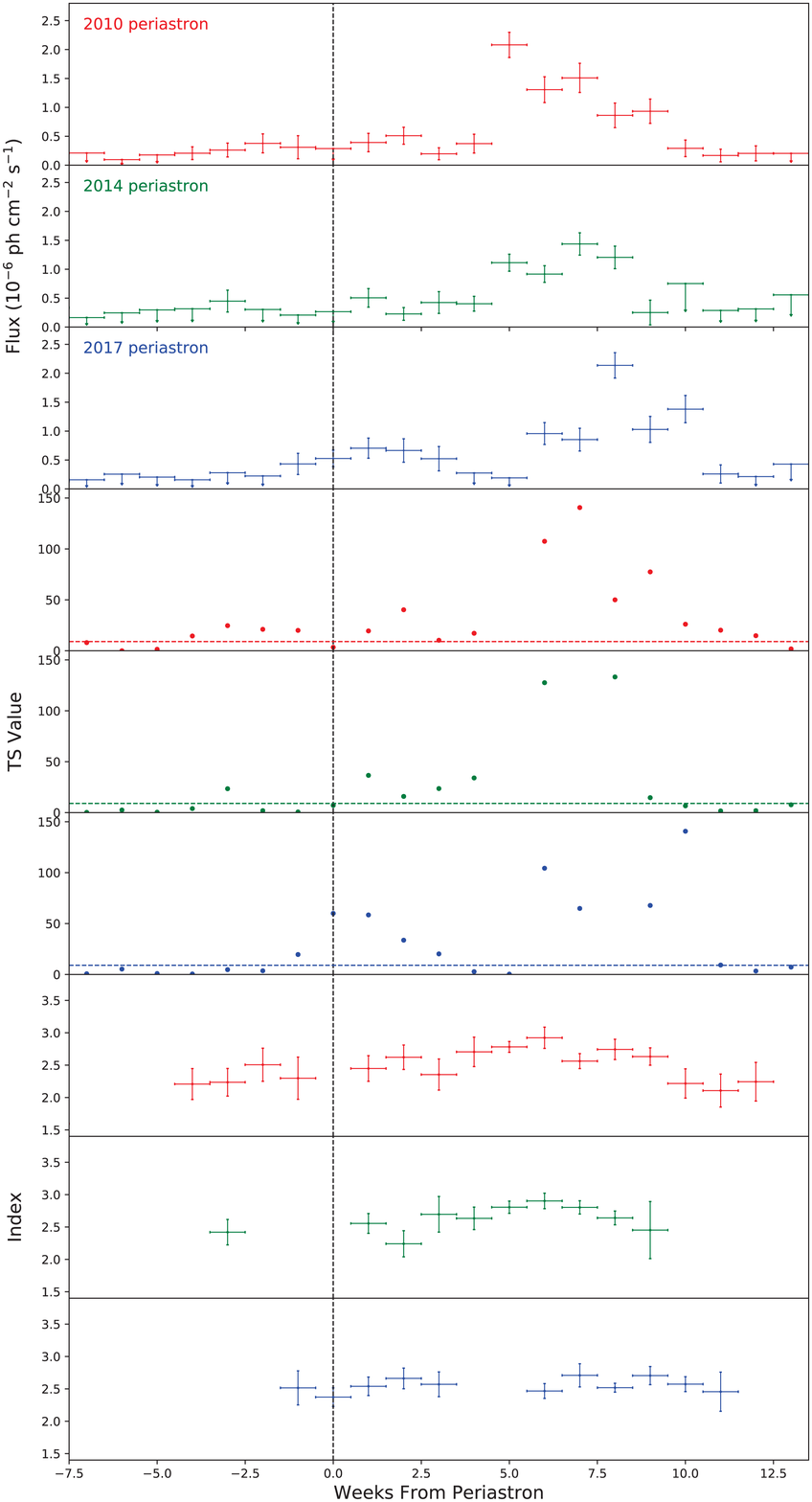}
  \includegraphics[width=0.48\textwidth] {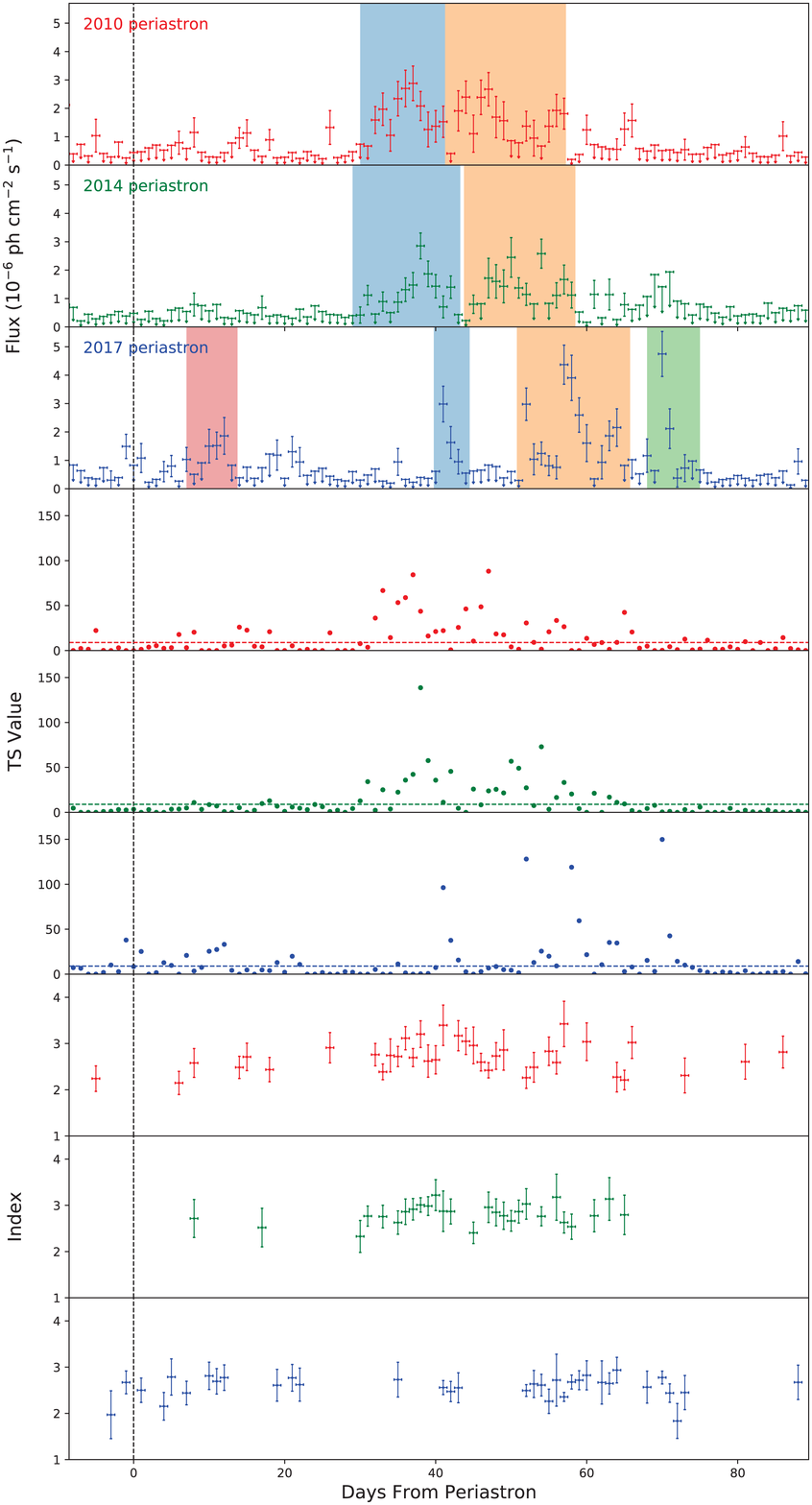}
  \caption{Weekly (left panels) and daily (right panels) 
    light curves (top three panels), TS values (middle three panels) and
    photon indexes (bottom three panels) 
    during \fermi flares as observed in 2010 (red), 2014 (green) and 2017 (blue).
    The vertical dashed black line in both panels indicates the time of periastron.}
  \label{fig_lc_daily_weekly}
\end{figure}

During PSR B1259-63's 2010 periastron passage \citep{abdo2011}
and the 2014 periastron passage \citep{caliandro2015},
bright GeV emission was observed $\sim$30--80\,d and $\sim$31--80\,d after the periastron point respectively.
For the 2017 periastron passage, the \fermi weekly and daily light curves are shown 
in Figure \ref{fig_lc_daily_weekly}(blue points in both panels).
A 95\% confidence upper limit was calculated when \psrb was not significantly detected
(Test Statistic (TS) value < 9)

It is apparent that a low flux level emission was detected around the 2017 periastron point,
similar to those reported in \cite{he2017}.
At about 10\,d after the periastron passage, an intense $\gamma$-ray flare became visible.
Then about 39\,d after the periastron point another flare with comparable flux showed up, 
followed by a series of additional flares.

For comparison with previous passages, 
we re-analyzed the Pass 8 data during the 2010 and 2014 periastrons with the latest IRFs.
The weekly and daily light curves are consistent with those reported in
\cite{abdo2011} and \cite{caliandro2015} respectively.
All the \fermi light curves of the periastron passages in 2010, 2014 and 2017
are shown in Figure \ref{fig_lc_daily_weekly}.

A clear similarity can be found in these three light curves.
They all have bright flares occurring at about one month after the periastron,
and there are one or more flares following the first one.
Apart from these similar trends,
clear differences are also obvious in the periastron light curves of the source
in 2010 2014 and 2017 in $\gamma$-rays.
Firstly, more flares are observed in 2017: one at time only 10\,d after the periaston passage,
and two more intense flares at time roughly 60\,d after the periastron passage.
In 2017, the most significant $\gamma$-ray flares were observed by \fermi at about 58\,d and 70\,d after the
periastron, but there were almost no visible $\gamma$-ray emissions in 2010 and 2014.
These results are consistent with \citet{johnson2017}.
The two flares at time around 30\,d after the periastron passage show a clear time delay
in each orbital period since 2010.

\begin{figure}
    \centering
    \includegraphics[width=0.96\textwidth] {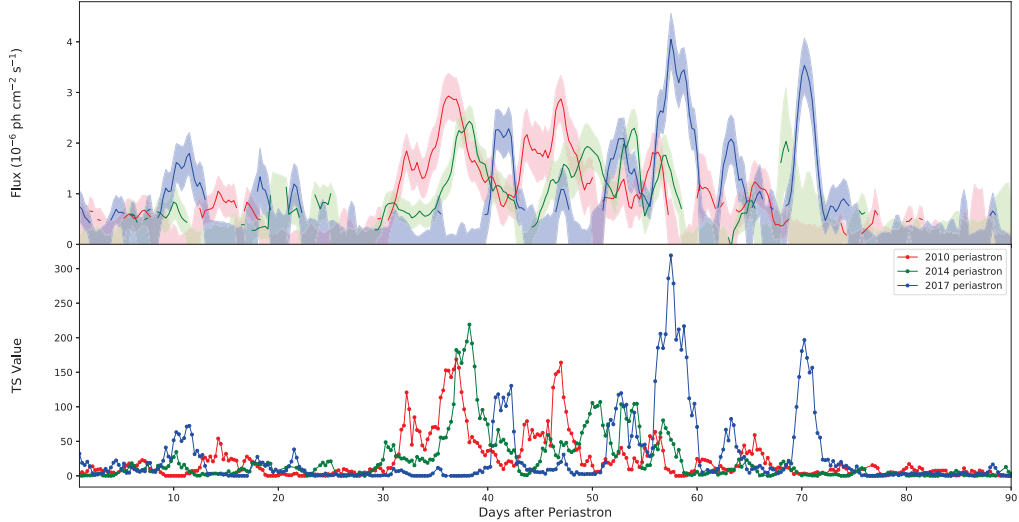}
    \caption{Sliding window light curves and TS values for flares observed by \fermi in  
      2010, 2014 and 2017 periastron passages of PSR B1259-63.
      The shadowed areas indicate the statistical error zones or the upper limits
        (at 95\% confidence level, when TS value is less than 9). 
      More details can be found in the text.}
    \label{fig_lc_sliding}
\end{figure}

In order to quantify the difference in the light curve profiles during these three periastron passages,
we produced smoothed light curves with the sliding windows technique 
introduced in \cite{caliandro2015}(Figure \ref{fig_lc_sliding}).
We choose the time windows of 2\,d, whose starting times lag behind the previous one by 6 hours.
During this analysis, a binned likelihood analysis was performed in every window.
The spectral index of PSR B1259-63 was allowed to vary between 1.0 and 4.0.
The smoothed light curves show the same trend as the daily light curve displayed in Figure \ref{fig_lc_daily_weekly},
but with the main structures standing out more clearly in the \fermi light curves.

To investigate the time lag of an individual $\gamma$-ray flare with respect to the one in the last orbital period, 
we did a cross correlation calculation using the results from the sliding windows technique.
We found that the 2014 flare is delayed $1.73\pm0.35$\,d with respect to the 2010 flare,
and the 2017 flare delayed $3.43\pm0.16$\,d with respect to the 2014 flare
within the time window of 30--55\,d associated with the perisatron passage,
where the two flares are always present in 2010, 2014 and 2017. 

\subsection{Spectral Analysis}

\begin{figure}
  \centering
  \includegraphics[width=0.48\textwidth] {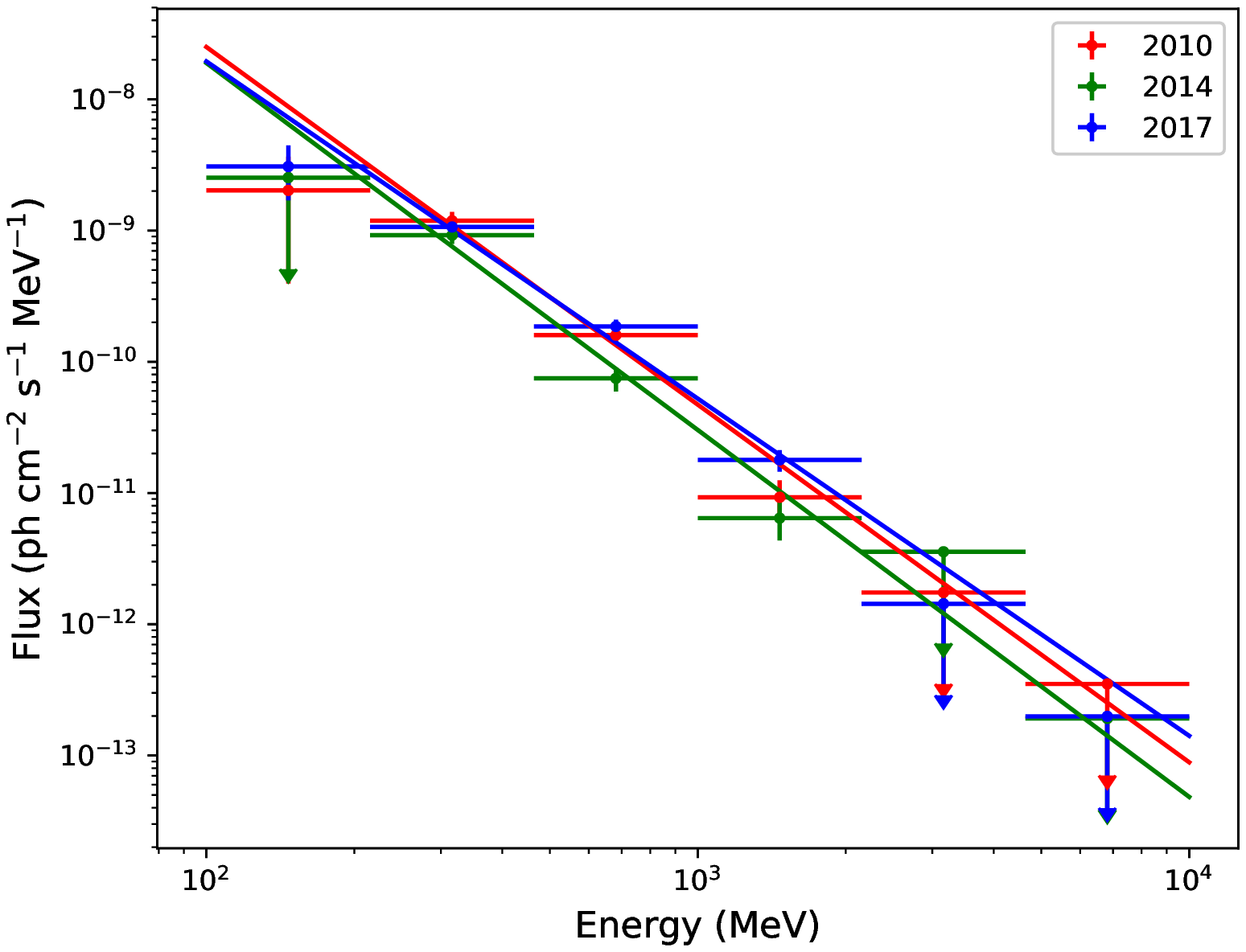}
  \includegraphics[width=0.48\textwidth] {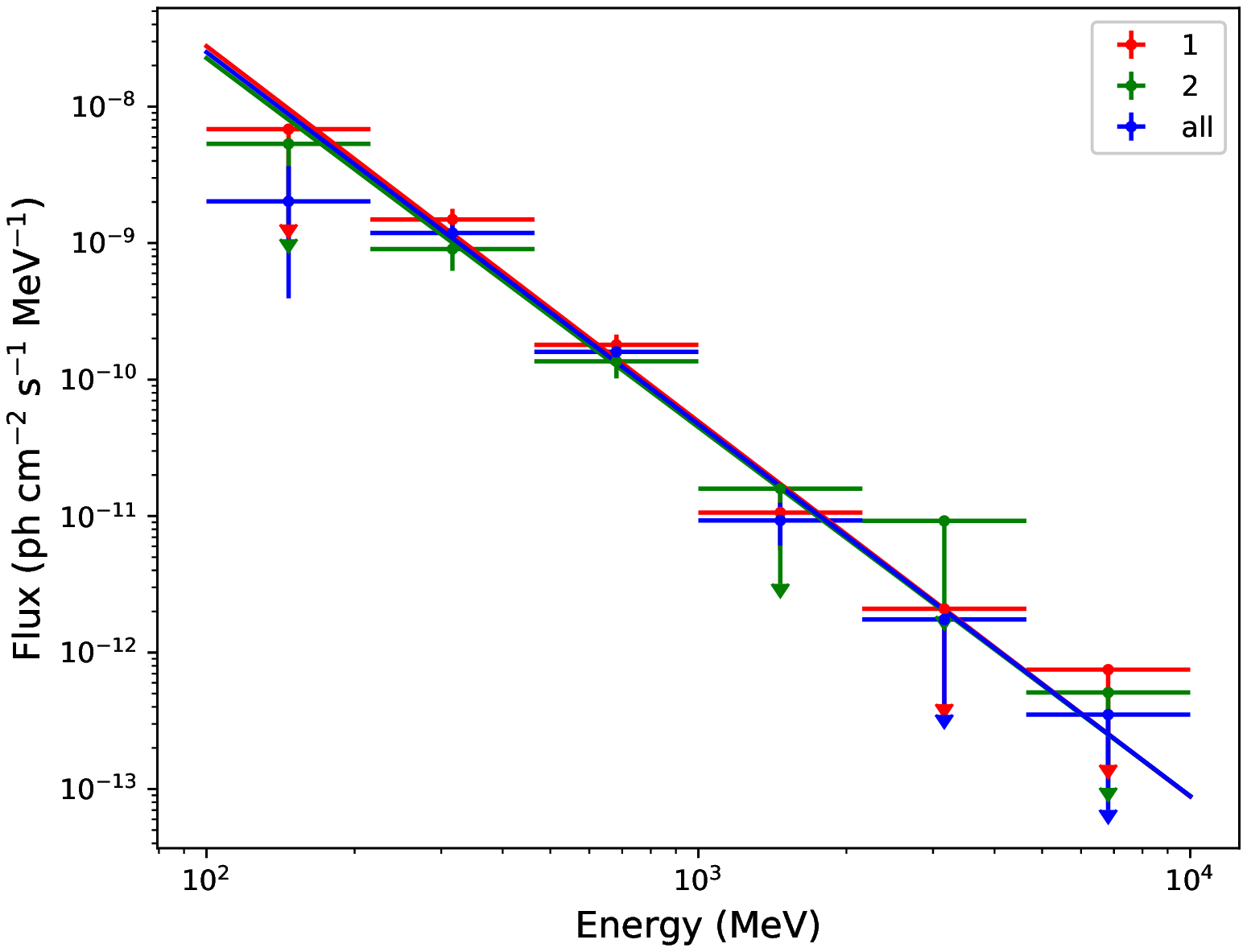}
  \includegraphics[width=0.48\textwidth] {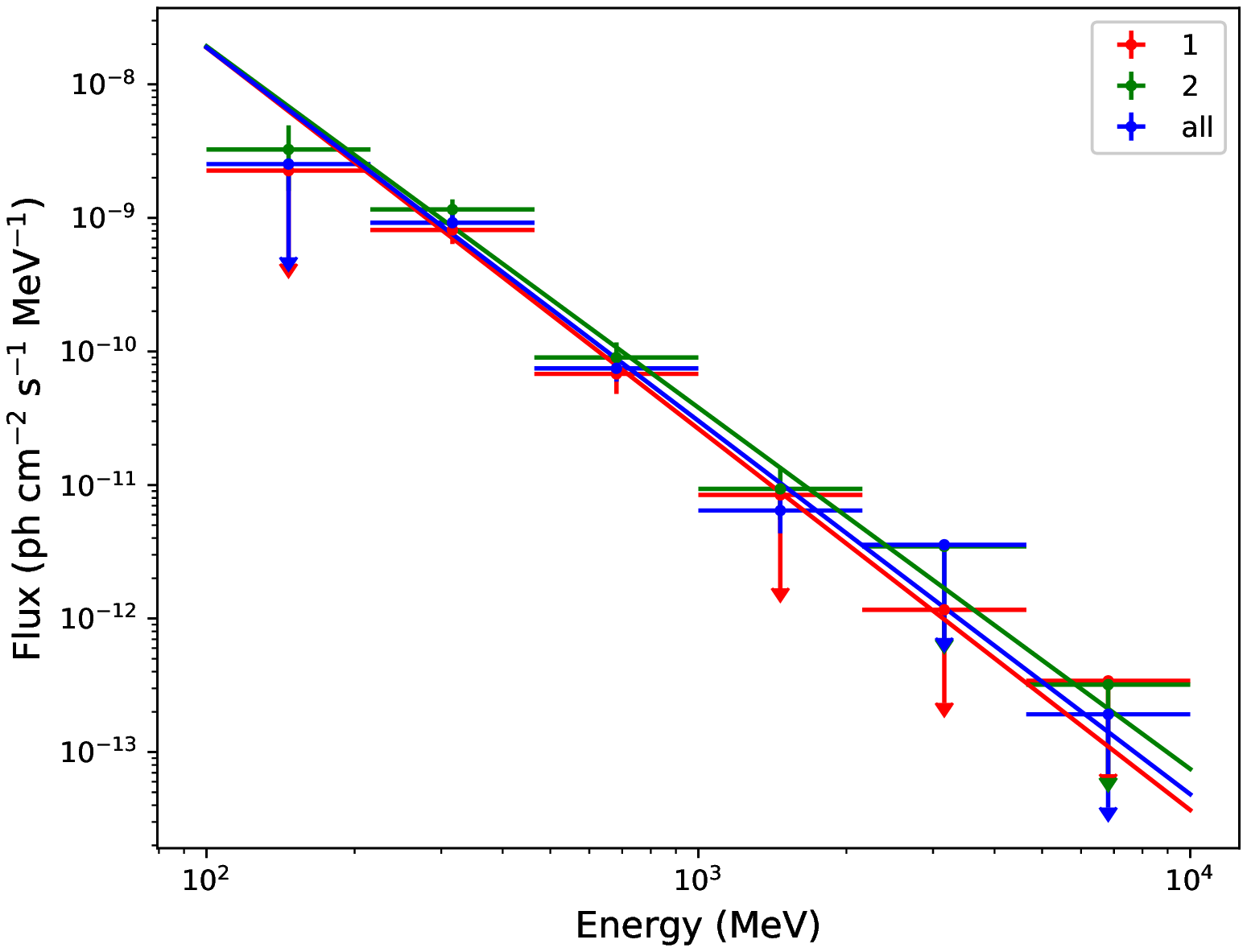}
  \includegraphics[width=0.48\textwidth] {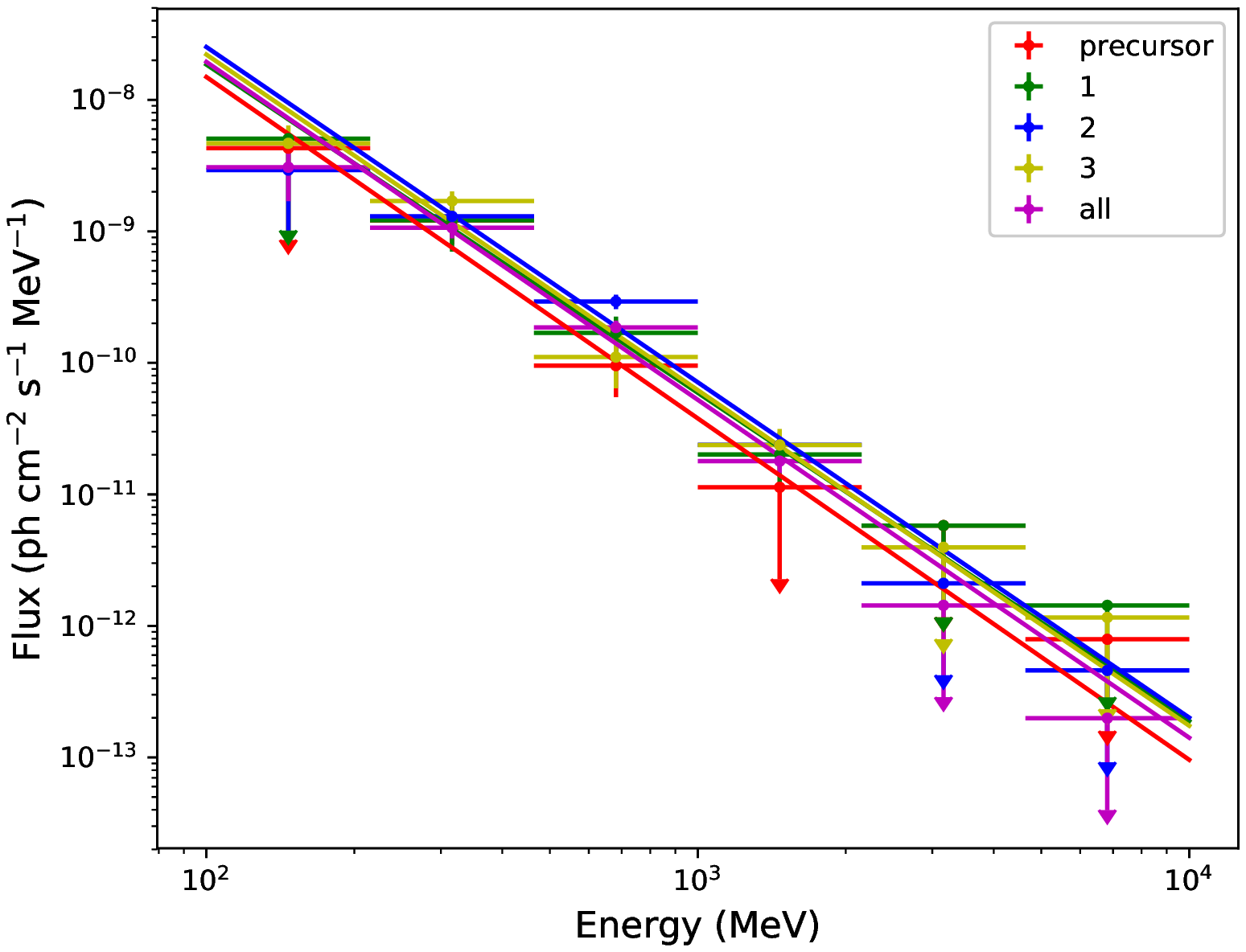}
  \caption{Spectrum of the flares observed by \fermi during the periastron periods of 2010, 2014 and 2017.
           The time intervals are defined in Table \ref{tab_time_int},
           which also presents the fitting parameters.
           $Top-left$: three periastrons with all data,
           $top-right$: each flare in the 2010 periastron,
           $bottom-left$: each flare in the 2014 periastron,
           and $bottom-right$: each flare in the 2017 periastron,}
  \label{fig_spec}
\end{figure}

To carry out spectral analysis of the 2010, 2014 and 2017 periastron passages,
we defined time intervals to denote the different flares.
As shown in Table \ref{tab_time_int} and Figure \ref{fig_lc_daily_weekly},
we define the flare periods as only positive \fermi detections (without upper limits)
in the continuous light curves in Figure \ref{fig_lc_sliding}.
The spectral parameters derived with standard likelihood analysis are also shown in Table \ref{tab_time_int}.

\begin{table}
    \begin{center}
        \caption{Spectral parameters of  flares from \psrb observed by \fermi 
        	    in the 2010, 2014 and 2017 periastron passages,
            fitted with a simple power-law model.
            The time intervals are labeled with differently colored shadows 
            in the right panels of Figure \ref{fig_lc_daily_weekly}.}
        \begin{tabular}{c|cccc}
            \hline
            Year & Flare & Time Interval    & Photon Index  & Flux (100\,MeV$\sim$10\,GeV)\\
                 &       &days after periastron & $\Gamma$  & 10$^{-6}$ ph cm$^{-2}$ s$^{-1}$ \\
            \hline
                 &  1    &  30$\sim$41.25   & 2.76$\pm$0.08 & 1.63$\pm$0.17 \\
            2010 &  2    & 41.25$\sim$57.25 & 2.72$\pm$0.08 & 1.40$\pm$0.16 \\
                 & all   &  30$\sim$57.25   & 2.74$\pm$0.05 & 1.51$\pm$0.12 \\
            \hline
                 &  1    &  29$\sim$43.25   & 2.85$\pm$0.07 & 1.06$\pm$0.10 \\
            2014 &  2    &  43.75$\sim$58.5 & 2.69$\pm$0.07 & 1.18$\pm$0.12 \\
                 & all   &  29$\sim$58.5    & 2.78$\pm$0.05 & 1.10$\pm$0.08 \\
            \hline
                 &precursor&7$\sim$13.75    & 2.63$\pm$0.12 & 0.99$\pm$0.21 \\
                 &  1    &  39.75$\sim$44.5 & 2.52$\pm$0.11 & 1.31$\pm$0.24 \\
            2017 &  2    & 50.75$\sim$65.75 & 2.57$\pm$0.06 & 1.69$\pm$0.15 \\
                 &  3    &  68$\sim$75      & 2.58$\pm$0.11 & 1.47$\pm$0.23 \\
                 & all   &  39.75$\sim$75   & 2.58$\pm$0.05 & 1.29$\pm$0.10 \\
            \hline
        \end{tabular}
        \label{tab_time_int}
    \end{center}
\end{table}

As shown in Figure \ref{fig_spec}, all the energy spectra can be represented with a simple power-law.
The spectral indices of the flares in 2010 and 2014 are comparable, but are softer than those in 2017.
The spectral index averaged over flares is about 2.74$\pm$0.05 in 2010,
2.78$\pm$0.05 in 2014 and 2.58$\pm$0.05 in 2017. 
It is obvious that the spectra became harder for flares in 2017.

\section{Discussion}
\label{sect:discussion}
We analyzed the \fermi data derived from observations of the periastron passages in 2010, 2014 and 2017.
We found that the energy spectrum from each flare can be represented with a simple power-law shape,
but with the spectral shape slightly changed.
The three $\gamma$-ray light curves indicate that,
in each flare there are two main peaks in 2010 and 2014, but 4 peaks in 2017.
The first main peaks of 2010 and 2014 are located at around 35\,d after the periastron passage,
and the two main peaks are delayed in 2014 by roughly 1.7\,d with respect to 2010.
In the 2017 flare, the source shows a precursor about 10\,d after the periastron passage,
but the following two peaks become weaker and lag behind those in 2014 by roughly 3.5\,d.
The strongest flares in 2017 occurred 58\,d and 70\,d after the periastron passage.  
 
It is generally thought that for a $\gamma$-ray binary system,
the high energy emission comes from the synchrotron or inverse Compton process,
in which particles are accelerated to relativistic speeds via shock 
between stellar wind of the companion star and pulsar wind from the neutron star.
In this scenario, the particles are accelerated in the shocked region,
which shows up with a bow shape and the power from the energetic pulsar wind will be partially extracted and
released in the form of high energy emissions. 
General support for this emission mechanism comes from 
the $\gamma$-ray binaries LS 5039 and LS I +61$^\circ$303,
for which soft $\gamma$-rays (hundreds of MeV to GeV) dominate the orbital phase region of the periastron,
where the adiabatic cooling prevents particles from being accelerated to high energies
\citep{takahashi2009, acciari2011, dubus2013, chang2016}.
While at apastron, this constraint becomes weaker 
so that the TeV and X-rays become dominant via inverse Compton and synchrotron processes.
Here a similar picture fits as well for \psrb in its two crossing points of the neutron star and the stellar disk:
the TeV and X-rays have emission peaks that are observed at the two crossing points.
The separation of the periastron point from the companion is much larger for \psrb 
than for LS 5039 and LS I +61$^\circ$303,
for example, the periastron distance of \psrb is 0.94\,AU,
which is much larger than that of 0.19\,AU for LS 5039 and
0.64\,AU for LS I +61$^\circ$303 in their apastron distance \citep{dubus2013}.
This in turn increases the difficulty if having \fermi flaring for \psrb
when it moves away from the stellar disk after the second crossing,
because the stellar density of the stellar disk is expected to become even smaller.
So far, almost all the theoretical models were developed to focus on the possible explanation of why
there is an emission peak at $\gamma$-rays as observed by \fermi.
Here the different timing properties as derived in this research based on \fermi observations definitely
put strong constraints upon the previous models, which, 
as is demonstrated in what follows, are hard to fully account for in these observational phenomena.

The first two observational results that 
any models have to explain are about the time delay after the periastron passage and
the total power output of the \fermi flare.
We know that the \fermi flare happens at time 30\,d later when the neutron star moves away from the stellar disk,
and the total power output of the flare is comparable to the rotational power of the pulsar.
To handle these, one needs either a condensed stellar environment in a shock model or a beaming effect in a jet model.   

\citet{chernyakova2015} proposed a destroyed stellar disk model to account for the time delay and
the large power output of the \fermi flare. 
Here they suggest that during the passage of the neutron star through the stellar disk,
the disk will be destroyed, probably due to the sound speed in the disk being lower than that of the neutron star.
As a result, part of the disk matter will accumulate and
surround the neutron star then the neutron star moves away from the stellar disk,
which provides a condensed stellar environment necessary for shock 
with the pulsar wind to accelerate particles to energies needed for having $\gamma$-ray emission.
In this scenario, the matter density of the shocked region may be comparable to those 
in periastron regions of LS 5039 and LS I +61$^\circ$303, 
where the MeV--GeV emissions are dominant. 
Since the neutron star is mostly enclosed within the accumulated matter, 
most power of the pulsar wind can be extracted via shock and emitted as $\gamma$-rays. 
It is obvious that this model expects only one flare at $\gamma$-rays 
once the neutron star passes through the stellar disk. 

Alternatively, an accretion/disk model was proposed recently \citep{yi2017}. 
Here they investigated the possibility of whether the stellar matter can form an accretion disk. 
They found that, under some conditions, the stellar matter can be captured in the vicinity of the neutron star. 
Since the angular momentum needs to be transported outward via viscosity,
an accretion disk can be formed a few days later once the neutron star passes through the stellar disk.

Then the disk will provide sufficient seed photons for inverse Compton scatterings off the relativistic pulsar wind.
The emission can be boosted to $\gamma$-rays via adjusting the Doppler factor to a proper value.
Again, in such an accretion/disk model,
the multiple peaks and time lag of the flare between orbits as observed by \fermi 
for the flares in 2010, 2014 and 2017 periastron passages are hard to properly handle. 

\citet{kong2012} considered the tail of the bow shock to understand the \fermi flare.
The collision between stellar wind and pulsar wind will form a bow-shaped shock.
Along with the evolution of the shock there will be two interesting regions:
one is the hot shock head where the normal high energy emissions are expected to be observed
in systems like LS 5039 and LS I +61$^\circ$303,
and another is the shock tail where the shocked matter will move all the way outwards while it is cooling off.
The entire bow shock has a cone shape and
its tail can pass through our line of the sight twice during movement of the neutron star in its orbit. 
The matter in the shock tail can have a moderate Doppler factor,
with which, because of inverse Compton scattering, 
the stellar photons can be boosted to $\gamma$-rays and show up as a \fermi flare twice. 
We see that in this scenario at least two main peaks as observed by \fermi can be predicted relatively naturally. 
However, additional features that were observed in the 2017 flare like the precursor, more flare structures
and the time lag of the flare with respect to the last orbital period remain hard to be properly addressed by the model.

Obviously, the emission properties of the $\gamma$-ray flares as observed by \fermi
in a time scope of 8 years which covers three orbital periastron passages are rather complicated.
A multi-wavelength observation campaign is needed 
to identify the emission mechanism left behind from these observational results, 
and distinguish the different models or disentangle degeneracy in model parameters 
which are so far only inferred from \fermi observations at $\gamma$-rays.  
In the case of jet models, if the inverse Compton peak is located at the MeV-GeV band, 
the synchrotron peak should be present at energies much lower than X-rays, most probably in the radio band. 
Therefore, a jet model may work if the radio peak can be detected in the SED by a sensitive radio telescope like FAST.
In the case where a shock model is applicable,
X-rays may be weak compared to those in the two crossing points of the stellar disk,
but could be expected to be visible by sensitive X-ray telescopes like $XMM-Newtion$ and
$NuSTAR$ at soft X-rays and/or $Insight-HXMT$ at hard X-rays after its next periastron passage, which is around 2020.
For the 2017 periastron passage, 
unfortunately the source was not observable by $Insight-HXMT$ because the solar avoidance angle was too small. 
A series of $Swift$/XRT observations was available in 2017 and
the data analysis/results will be reported in a forthcoming paper (Chang et al. 2018, in preparation).

For the three periastron passages we analyzed in this paper, the times of the main flares are consecutively delayed.
It is possible that this is due to variation of the Be stellar disk in \psrb.
Quasi-periodic variation of the Be stellar disk was also observed 
in LS I +61$^\circ$303 \citep{z1999,z2000}, 
which lead to multiwavelength super-orbital modulation
\citep{gregory2002,li2012,li2014,ackermann2013,ahnen2016}
A similar super-orbital modulation may also exist in \psrb, 
but it is difficult to confirm considering the long orbital period of 3.4 years:
future long-term monitoring of \psrb may shed light on this.


\begin{acknowledgements}
We acknowledge the suggestion, discussion and help from Dr. Jian Li in this research.
The authors thank support from the National Key R\&D Program of China (grant No. 2016YFA0400800), 
the NSFC (Nos. U1838201, U1838202, 11733009 and 11473027), XTP project (XDA 04060604)
and the Strategic Priority Research Programme `The Emergence of Cosmological Structures'
of the Chinese Academy of Sciences (Grant No.XDB09000000).
\end{acknowledgements}

\label{lastpage}

\end{document}